\documentclass[twocolumn,aps,prb,showpacs,groupedaddress]{revtex4-1}

\usepackage[latin9]{inputenc}
\usepackage{float}
\usepackage{color}
\usepackage{bm}
\usepackage{amsmath}
\usepackage{amssymb}
\usepackage{graphicx}
\usepackage{epstopdf}
\makeatletter

\@ifundefined{textcolor}{}
{%
 \definecolor{BLACK}{gray}{0}
 \definecolor{WHITE}{gray}{1}
 \definecolor{RED}{rgb}{1,0,0}
 \definecolor{GREEN}{rgb}{0,1,0}
 \definecolor{BLUE}{rgb}{0,0,1}
 \definecolor{CYAN}{cmyk}{1,0,0,0}
 \definecolor{MAGENTA}{cmyk}{0,1,0,0}
 \definecolor{YELLOW}{cmyk}{0,0,1,0}
}

\usepackage{latexsym}
\usepackage[ngerman, english]{babel}
\usepackage{tikz}
\makeatother

\newcommand{\abs}[1]{\left\lvert#1\right\rvert}

\begin{document}

\title{Thermodynamics of frustrated ferromagnetic spin-$1/2$ Heisenberg
chains: The role of inter-chain coupling}

\author{P. M\"uller}

\author{J. Richter}

\affiliation{Institut f\"ur Theoretische Physik,
Otto-von-Guericke-Universit\"at Magdeburg,
D-39016 Magdeburg, Germany}

\author{D. Ihle}

\affiliation{Institut f\"ur Theoretische Physik, Universit\"at Leipzig, D-04109 Leipzig,
Germany}

\date{\today}
\begin{abstract}
The thermodynamics of coupled frustrated ferromagnetic chains is studied
within a spin-rotation-invariant Green's function approach.  We consider an
isotropic Heisenberg
spin-half system
with a ferromagnetic in-chain coupling $J_1<0$ between nearest neighbors and a
frustrating antiferromagnetic
next-nearest neighbor in-chain coupling $J_2>0$.  We focus on moderate
strength of frustration $J_2 <
|J_1|/4$
such that the in-chain spin-spin correlations are predominantly ferromagnetic.
We consider two inter-chain couplings (ICs) $J_{\perp,y}$ and $J_{\perp,z}$,
corresponding to the two axis perpendicular to the chain, where
ferromagnetic as
well as antiferromagnetic ICs are taken into account.
We discuss the influence of frustration on the ground-state properties for
antiferromagnetic ICs, where the ground state is of quantum nature.
The major part of our study is devoted to the finite-temperature properties.  
We calculate the critical temperature $T_{c}$
as a function of the competing exchange couplings $J_{2},J_{\perp,y},
J_{\perp,z}$.  
We find that for fixed ICs $T_c$ monotonically decreases with increasing
frustration $J_2$, where as $J_2 \to |J_1|/4$ the $T_c(J_2)$-curve drops
down rapidly.  To characterize the magnetic ordering below and above $T_c$
we calculate the spin-spin correlation functions $\langle {\bf S}_0 {\bf
S}_{\bf R} \rangle$, the magnetic order parameter $M$, 
the uniform static susceptibility $\chi_0$ as well as the correlation length
$\xi$.  Moreover, we discuss the specific heat $C_V$ and the temperature
dependence of the excitation spectrum $\omega_{\mathbf{q}}$.  As $J_2 \to
|J_1|/4$ some unusual frustration-induced features were found, such as 
an increase of the in-chain spin stiffness (in case of ferromagnetic ICs) or
of the in-chain spin-wave velocity (in case of antiferromagnetic ICs) with
growing temperature.  \end{abstract}
\maketitle

\section{Introduction\protect \\
}
                              \label{sec:intro}

One-dimensional (1D) frustrated quantum $J_1$-$J_2$ Heisenberg systems have been studied intensively
for many
years.\cite{bader,hamada,DKO97,theo1,krivnov2007,theo2,theo3,theo4,theo5,theo6,theo7,theo8,theo9,
Vekua2007,Momoi2007,Sudan2009,theo11,theo13,theo14,theo15,theo16,theo17,theo18,zinke2009,theo19,theo20,
RGMchainfrusferro,RGMchainfrusferromagnet,RGMchainarbitraryspin} 
They exhibit a large variety of physical many-body phenomena. 
Many experimental studies have shown that there is a 
plethora of materials, such as the edge-shared cuprates
LiVCuO$_4$, LiCu$_2$O$_2$, NaCu$_2$O$_2$,
Li$_2$ZrCuO$_4$, Ca$_2$Y$_2$Cu$_5$O$_{10}$, and Li$_2$CuO$_2$,
which can be adequately described by a chain model with ferromagnetic (FM) nearest neighbors
(NN) interaction $J_1$ and antiferromagnetic (AFM)
next-nearest neighbors (NNN) interaction $J_2$. 
\cite{gippius2004,enderle2005,drechsler2007,drechsler2007a,buettgen2007,fong99,matsuda99,rosner,rosner2,exp1,exp2,exp3,exp4,exp5,exp6,exp7,exp9,exp10,exp11}

From the experimental point of view it is clear that an inter-chain coupling (IC)
is unavoidably present in real materials, that leads to three-dimensional (3D)
physics at least at low temperatures, and, in particular, it may lead to a phase
transition to a magnetically long-range ordered phase below a critical
temperature $T_c$. Thus, for example,   in
Refs.~\onlinecite{fong99,matsuda99,rosner2}  for the magnetic-chain
material Ca$_2$Y$_2$Cu$_5$O$_{10}$ the following parameters were reported  $J_1\approx-93$ K (FM), $J_2\approx4.7$ K (AFM),
and $T_c\approx30$ K, indicating the presence of a non-negligible IC. 
The discussion of the role of the IC makes the theoretical treatment more
challenging, since several tools, such as the Density-Matrix Renormalization Group (DMRG) and the Exact Diagonalization (ED), are less effective in
dimension $D>1$.  
In fact, coupled frustrated spin-chains are much less investigated in
literature.
Moreover, most of these investigations 
were focused on ground state (GS)
properties.\cite{theo6,zinke2009,Ueda2009,prl2011,theo3,prb2015,Du2016}

In our paper we want to discuss the role of 
the IC in coupled frustrated spin-$1/2$ chain magnets 
with a FM NN in-chain  
coupling $J_1<0$ and an AFM NNN in-chain coupling
$J_2>0$.
According to Fig.~\ref{fig1} the chains are aligned along the $x$-axis, and
they
are coupled along the $y$- and
$z$-axis by $J_{\perp,y}$ and $J_{\perp,z}$, respectively.
The two NN ICs $J_{\perp,y}$ and $J_{\perp,z}$ are treated
as independent variables which can be  FM as well as AFM.
The corresponding model reads
\begin{eqnarray} 
H & = & J_{1}\sum_{\langle i,j\rangle,x}\bm{S}_{i}\cdot\bm{S}_{j}
+J_{2}\sum_{\left[i,j\right],x}\bm{S}_{i}\cdot\bm{S}_{j} \;
\nonumber \\ 
  & + & J_{\perp,y}\sum_{\langle i,j\rangle,y}\bm{S}_{i}\cdot\bm{S}_{j}+J_{\perp,z}\sum_{\langle
 i,j\rangle,z}\bm{S}_{i}\cdot\bm{S}_{j}, \label{eq_ham}
\end{eqnarray}
where $\langle i,j\rangle, x,y,z$ labels NN bonds along the
corresponding axis and $\left[i,j\right],x$ labels NNN bonds  along the
chain.  Moreover, we consider $J_1 < 0$ and  $J_2 \ge 0$, whereas no sign
restrictions are valid for $J_{\perp,y}$ and
$J_{\perp,z}$.

An appropriate method to study thermodynamic properties of the model
(\ref{eq_ham}) in the whole
temperature range is the second-order rotation-invariant
Green's function method, see, e.g., 
Refs.~{\onlinecite{kondoyamaji,RhoScal94,ShiTak1991,SSI94,barabanov94,winterfeldt97,ihle2001,
canals2002,prb2004,RGMchainfrusferro,RGM2Dj1j2frusferro,RGMSgg1anisotropy,RGMquasi2Dvskagomelattice,
schmal2006,RGMchainfrusferromagnet,RGMmoritzcollinearstripe,bcc_rgm_j1j2,RGMchainarbitraryspin}.
This method has been used recently for the 1D $J_1$-$J_2$
model,\cite{RGMchainfrusferro,RGMchainarbitraryspin} for the
frustrated square-lattice ferromagnet\cite{RGM2Dj1j2frusferro}  as well as for
the 3D frustrated ferromagnet on the body-centered cubic lattice.
\cite{bcc_rgm_j1j2}
%}

\begin{figure}
%\centering \includegraphics[scale=1]{New_Tikzfile.eps}
\centering \includegraphics[scale=1]{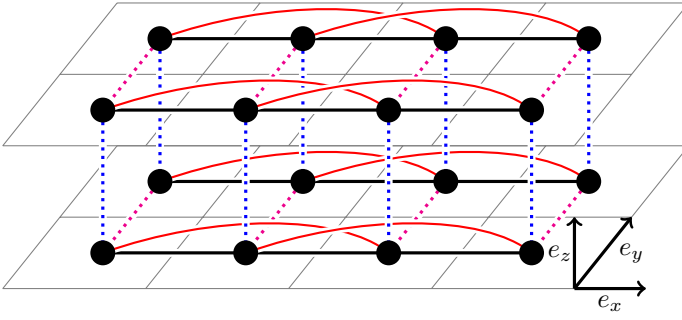}
\protect\caption{(Color online) Sketch of the considered model of coupled frustrated spin chains: $J_1$ - NN in-chain coupling
(solid black),
$J_2$ -- NNN in-chain coupling (solid red), 
$J_{\perp,y}$ -- NN inter-chain coupling in $y$-direction (dotted magenta),  $J_{\perp,z}$
-- NN inter-chain coupling
in $z$-direction  (dotted blue).
\label{fig1} }
\end{figure}
For the classical model (\ref{eq_ham}) in $D=1$ (i.e., $s\rightarrow\infty$
and
$J_{\perp,y}=J_{\perp,z}=0$)
the critical strength of frustration, where the FM GS breaks
down, is $J_{2}^{c,clas}=|J_1|/4$, 
which is also the quantum-critical point $J_2^{c}$ for the spin-$1/2$ model.
\cite{bader}
For $J_2 <  J_2^{c}$ the GS is FM, whereas for
$J_2 > J_2^{c}$ the GS is a quantum spin singlet with incommensurate spiral
correlations. \cite{bader,hamada,DKO97,krivnov2007}
On the classical level, the spiral phase does not depend on the IC
couplings $J_{\perp,y}$, or $J_{\perp,z}$ respectively, whereas for the
quantum model the  spiral phase does depend on the IC coupling,   
see, e.g., Refs.~\onlinecite{zinke2009} and \onlinecite{Du2016}.

In the
present paper we will focus  on the parameter region of weak frustration $J_2< J^c_2$.
Although, for those values of $J_2$ 
the GS is FM (i.e. it is a classical state without
quantum fluctuations),
the frustrating
NNN bond $J_2$ may influence the thermodynamics  substantially, in particular
in the vicinity
of the zero-temperature
transition, i.e., at $J_2 \lesssim J^c_2$. 
\cite{RGMchainfrusferro,RGMchainarbitraryspin,RGM2Dj1j2frusferro,bcc_rgm_j1j2,Katanin2}

We mention here that the case of coupled  AFM
spin-$1/2$ Heisenberg
chains is well studied, see, e.g.,
Refs.~\onlinecite{HJSchulz,Irkhin,Bocquet,Zvyagin}.
Since in this case the GS of the isolated chain is of quantum nature and
does not exhibit magnetic long-range order the behavior for small IC is
different to our case of FM chains.   

It is appropriate to notice that in real edge-shared cuprates
often the  inter-chain coupling  is
more sophisticated than that we consider in our paper. Moreover, there is a
large variety in the topology of the IC, see, e.g.,
Ref.~\onlinecite{prb2015}.
However, the simplest case of a perpendicular IC $J_\perp$ 
corresponds, e.g.,
to LiVCuO$_4$ and
Li(Na)Cu$_2$O$_2$.\cite{gippius2004,rosner,enderle2005,buettgen2007}
Furthermore, we note that most of these compounds exhibit spiral spin-spin
correlations along the chain direction, i.e., the frustration exceeds 
$J^c_2$.
Hence, there is no direct relation of our results to those compounds
with $J_2> J^c_2$,
and the focus here is on 
the general
question for the crossover from a purely 1D $J_1-J_2$ ferromagnet to a
quasi-1D and finally to a 3D system.

\section{Rotation-invariant Green's function method (RGM)}
\label{methods}

The RGM has been widely applied to frustrated quantum
spin systems. \cite{RGMchainfrusferro,RGMchainarbitraryspin,RGMchainfrusferromagnet,barabanov94,ihle2001,canals2002,prb2004,schmal2006,RGMmoritzcollinearstripe,RGM2Dj1j2frusferro,bcc_rgm_j1j2}
Therefore, we illustrate here only some basic relevant features of
the method. At that we follow
Refs.~\onlinecite{RGMchainfrusferro} and \onlinecite{bcc_rgm_j1j2}.
The retarded two-time Green's function in momentum space $\langle\langle S_{\mathbf{q}}^{+};S_{-\mathbf{q}}^{-}\rangle\rangle_{\omega}=-\chi_{\mathbf{q}}^{+-}(\omega)$ 
determines the spin-spin correlation functions and the thermodynamic quantities.
The equation of motion in the second order using spin rotational symmetry, i.e., $\langle S_{i}^{z}\rangle=0$, is
expressed as $\omega^{2}\langle\langle S_{\mathbf{q}}^{+};S_{-\mathbf{q}}^{-}\rangle\rangle_{\omega}=
M_{\mathbf{q}}+\langle\langle-\ddot{S}_{\mathbf{q}}^{+};S_{-\mathbf{q}}^{-}\rangle\rangle_{\omega}$
with $M_{\mathbf{q}}=\left\langle \left[[S_{\mathbf{q}}^{+},H],S_{-\mathbf{q}}^{-}\right]\right\rangle $
and $-\ddot{S}_{\mathbf{q}}^{+}=\left[[S_{\mathbf{q}}^{+},H],H\right]$.
For our model (\ref{eq_ham}) the moment $M_{\bf{q}}$ is given by
\begin{eqnarray}
M_{\bf{q}}&=&4 J_{1} c_{100} (\cos (q_x)-1)+4 J_{2} c_{200} (\cos (2 q_x)-1)\\ \label{eq_moment}
&+& 4 J_{\perp,y} c_{010} (\cos (q_y)-1)+4 J_{\perp,z} c_{001} (\cos (q_z)-1), \nonumber
 \end{eqnarray}
where $c_{hkl}\equiv c_{\bm{R}}=\langle S_{\bm 0}^{+}S_{\bm{R}}^{-}\rangle=2\langle
{\bf S}_{\bm 0}{\bf S}_{\bm{R}}\rangle/3$,
$\bm{R}=h\bm{a}_{1}+k\bm{a}_{2}+l\bm{a}_{3}$, ($\bm{a}_{j}$ are the cartesian unit vectors).
For the second derivative $-\ddot{S}_{i}^{+}$ we apply the decoupling scheme in real
space \cite{kondoyamaji,ShiTak1991,RhoScal94,SSI94,barabanov94,winterfeldt97,ihle2001}
\begin{align}\label{eq:entk1}
S_{i}^{+}S_{j}^{+}S_{k}^{-} & =\alpha_{i,k}\langle S_{i}^{+}S_{k}^{-}\rangle 
S_{j}^{+}+\alpha_{j,k}\langle S_{j}^{+}S_{k}^{-}\rangle S_{i}^{+}, 
\end{align}
where  $i\neq j\neq k\neq i$ and the quantities $\alpha_{i,j}$ are vertex parameters introduced to improve the
decoupling approximation.
In the
minimal version of the RGM we consider as many vertex parameters as
independent conditions for them can be found, i.e., we have
$\alpha_{x}$, $\alpha_{y}$, and $\alpha_{z}$,  related to in-chain
($\alpha_{x}$) and inter-chain correlators ($\alpha_{y}$ and $\alpha_{z}$).

By using the operator identity $\bm{S}_{i}^{2} 
=S_{i}^{+}S_{i}^{-}-S_{i}^{z}+(S_{i}^{z})^{2}$ we get the sum rule
\begin{align} \label{eq:op_ident}
\langle S_{j}^{-}S_{j}^{+}\rangle=\langle S_{j}^{+}S_{j}^{-}\rangle=\frac{1}{2},
\end{align}
where $\langle S_j^z\rangle=0$ was used.
The decoupling scheme (\ref{eq:entk1}) leads to the equation 
$-\ddot{S}_{\mathbf{q}}^{+}=\omega_\mathbf{q}^2{S}_{\mathbf{q}}^{+}$ in momentum
space.
Then we get 
\begin{equation}
\chi^{+-}_{\mathbf{q}}(\omega) =- \langle\langle S_{\mathbf{q}}^{+};S_{-\mathbf{q}}^{-}\rangle\rangle_{\omega}
=\frac{M_{\mathbf{q}}}{\omega_{\mathbf{q}}^2-\omega^2}
\end{equation}
with the dispersion relation
\begin{eqnarray}
\omega_{\bf{q}}^{2} & = & \sum_{n}J_{n}^{2}(1-\text{cos}(\mathbf{r}_{n}\mathbf{q}))(1+2p_{2\mathbf{r}_{n}}-2p_{\mathbf{r}_{n}})\nonumber \\
 & - & \sum_{n}J_{n}^{2}(1-\text{cos}(\mathbf{r}_{n}\mathbf{q}))(4\text{cos}(\mathbf{r}_{n}\mathbf{q})p_{\mathbf{r}_{n}})
 \label{omega}\\
 & + & \sum_{n\neq m}J_{n}J_{m}(1-\textrm{cos}(\mathbf{r}_{n}\mathbf{q}))\left(4p_{\mathbf{r}_{n}+\mathbf{r}_{m}}
    -4\textrm{cos}(\mathbf{r}_{m}\mathbf{q})p_{\mathbf{r}_{n}}\right)\nonumber \\
 & + & 2J_{1}J_{2}\left(1-\textrm{cos}(q_{x})\right)\left(3+2\textrm{cos}(q_{x})\right)\left(p_{(1,0,0)}-p_{(3,0,0)}\right),\nonumber 
 \end{eqnarray}
where 
the following abbreviations are used: 
\begin{eqnarray}
 J_{3} =  J_{\perp,y},  \; \;  J_{4}  = J_{\perp,z}, \hspace{5.1cm} \nonumber \\
 r_{1}  =  (1,0,0), \; 
 r_{2}  = (2,0,0), \;  
 r_{3}  = (0,1,0), \;  
 r_{4}  = (0,0,1), \nonumber \\
 p_{(n,0,0)}=\alpha_x c_{n00},  \;  \;
 p_{(m,n,0)}=\alpha_y c_{mn0}, \hspace{2.7cm} \nonumber \\
 p_{(m,0,n)}=\alpha_z c_{m0n}, \;  \;
 p_{(0,n,m)}=(\alpha_y+\alpha_z)c_{0nm}/2 \;.\hspace{1.12cm}
\end{eqnarray}
Moreover, 
lattice symmetry is exploited to reduce the number of non-equivalent
correlators entering Eq.~(\ref{omega}). 
Expanding $\omega_{\bf{q}}$
around
$\mathbf{q} = \Gamma=(0,0,0)$ we find
 $\frac{\partial\omega_{\mathbf{q}}}{\partial
q_i}\vert_{\mathbf{q}=\mathbf{0}}
= v_i$ and 
 $\frac{\partial^2\omega_{\mathbf{q}}}{2\partial q_i^2}\vert_{\mathbf{q}=\mathbf{0}}= \rho_i$.
Here the quantities $v_i$, $i=x,y,z$, are the spin-wave velocities relevant for AFM
$J_\perp$, and $\rho_i$, $i=x,y,z$, are the spin-stiffness parameters relevant for FM
$J_\perp$.
The corresponding
equations for the spin-wave velocities $v_i$ (Eqs.~(\ref{velocity_x}),(\ref{velocity_y}) and (\ref{velocity_z}))
and for the spin stiffnesses $\rho_i$ (Eqs.~(\ref{stiffness_x}),(\ref{stiffness_y}) and (\ref{stiffness_z})) 
are provided in the Appendix.

The uniform static spin susceptibility is obtained via $\chi_0 =
\lim_{{\mathbf{q}} \to {\mathbf{0}}}\chi_{\bf{q}}$,
$\chi_{\bf{q}}=\chi_{\bf{q}}(\omega=0)=\chi^{+-}_{\bf{q}}(\omega=0)/2$.
The explicit expression for 
$\chi_0$ is given in the Appendix, see, Eqs.~(\ref{eq_chi0}), (\ref{eq_chi01}) (\ref{eq_chi02}), and 
(\ref{eq_Delta3}). (Note
that finally Eqs.~(\ref{eq_chi0}), (\ref{eq_chi01}) and  (\ref{eq_chi02})
yield $\chi_0=\chi_0^{(1)}=\chi_0^{(2)}=\chi_0^{(3)}$, because of
the 
isotropy constraint, see below.)
The correlation functions $c_{\bm{R}}=\frac{1}{N}\sum_{\mathbf{q}}c_{\mathbf{q}}\text{e}^{i\bm{qR}}$
are given by the spectral theorem,\cite{Tya67} 
\begin{equation}
c_{\mathbf{q}}=\langle S_{\mathbf{q}}^{+}S_{-\mathbf{q}}^{-}\rangle=\frac{M_{\mathbf{q}}}
{2\omega_{\mathbf{q}}}[1+2n(\omega_{\mathbf{q}})],\label{eq_C_q}
\end{equation}
where $n(\omega)=(\text{e}^{\omega/T}-1)^{-1}$ is the Bose-Einstein
distribution function. 
In the long-range ordered phase the correlation function $c_{\bm{R}}$
is written as \cite{ShiTak1991,winterfeldt97,junger2009,RGMmoritzcollinearstripe}
\begin{equation}
c_{\bm{R}}=\frac{1}{N}\sum_{\mathbf{q}\ne\mathbf{Q}}c_{\mathbf{q}}\text{e}^{i\bm{qR}}+\text{e}^{i\bm{QR}}C_{\mathbf{Q}},
\end{equation}
where $c_{\mathbf{q}}$ is given by Eq.~(\ref{eq_C_q}). 
The condensation term $C_{\mathbf{Q}}$, i.e. the long-range part of the correlation functions, is associated with the magnetic wave vector $\mathbf{Q}$, 
which describes the magnetically long-range ordered
phase.
 Depending
on the sign of $J_{\perp,y}$ and $J_{\perp,z}$ the
 magnetic wave vector 
is $\mathbf{Q}=(0,Q_y,Q_z)$, where $Q_y=0$  ($Q_z=0$) for FM $J_{\perp,y}<0$
($J_{\perp,z}<0$) and  $Q_y=\pi$  ($Q_z=\pi$) for AFM $J_{\perp,y}>0$ 
($J_{\perp,z}>0$). 
The order parameter, i.e. the corresponding (sublattice) magnetization $M$,
is connected with the condensation term by the formula $M=\sqrt{3C_\mathbf{Q}/2}$.
The magnetic correlation length $\xi_\mathbf{Q}$ in the paramagnetic regime ($T>T_c$) is
obtained by expanding the static susceptibility $\chi_\mathbf{q}$
around the magnetic wave-vector $\mathbf{Q}$, i.e. $\chi_\mathbf{q}\sim\chi_\mathbf{Q}/(1+\xi_\mathbf{Q}^2(\mathbf{Q-q})^2)$, 
see, e.g.,
Refs.~\onlinecite{winterfeldt97,RGMquasi2Dvskagomelattice,RGMSgg1anisotropy,RGMmoritzcollinearstripe}.

Finally we have to make sure that as many equations are provided as unknown quantities
are given. Obviously the inverse Fourier transformation of Eq.~(\ref{eq_C_q}) yields an equation for each spatial
spin-spin correlation function appearing in the system of coupled equations
that has to be solved numerically.
Three more equations are required to determine the vertex parameters $\alpha_x$, $\alpha_y$ and $\alpha_z$.
One equation is provided by the sum rule
Eq.~(\ref{eq:op_ident}), and the remaining 
two equations are obtained by the isotropy constraint, see , e.g., 
Refs.~\onlinecite{RGMquasi2Dvskagomelattice,RGMSgg1anisotropy,RGMmoritzcollinearstripe}, 
i.e. the static susceptibility $\chi_{\mathbf{q}}$ has to be isotropic in the limit
$\mathbf{q}\rightarrow\mathbf{0}$:
$\lim_{q_z\to0}\chi(q_x=0,q_y=0,q_z)=\chi_0^{(1)}=\lim_{q_y\to0}\chi(q_x=0,q_y,q_z=0)=\chi_0^{(2)}$ 
and
$\lim_{q_z\to0}\chi(q_x=0,q_y=0,q_z)=\lim_{q_x\to0}\chi(q_x,q_y=0,q_z=0)=\chi_0^{(3)}$,
where 
analytical expressions for $\chi_0^{(i)}$, $i=1,2,3$, are given in the
Appendix, see Eqs.~(\ref{eq_chi0}) - (\ref{eq_Delta3}).
Moreover, in the magnetically ordered phase we use 
the divergence of the static susceptibility $\chi^{-1}_{\bf{Q}}=0$ at the corresponding magnetic
wave-vector $\mathbf{Q}$  to calculate the condensation term $C_{\bf{Q}}$, see e.g. Refs. 
\onlinecite{junger2009,RGMmoritzcollinearstripe,bcc_rgm_j1j2}.
For antiferromagnetic IC ($J_{\perp,y}>0$ and $J_{\perp,z}>0$), for instance,
the relevant staggered susceptibility $\chi_{(0,\pi,\pi)}$ is given by
Eq.~(\ref{eq_chi0pp}), and the condition for long-range order reads as
$\Delta_{(0,\pi,\pi)}=0$, see Eq.~(\ref{eq_delta0pp}), which corresponds to the
vanishing of the gap in $\omega_{\mathbf{q}}$ at
${\mathbf{q}}={\mathbf{Q}}={(0,\pi,\pi)}$.

\section{Results}
 Although,
the two ICs $J_{\perp,y}$ and $J_{\perp,z}$ are treated
as independent variables in our theory, in what follows we will consider  
the case with identical ICs in $y$- and
$z$-direction, i.e. $J_{\perp,y}=J_{\perp,z}=J_{\perp}$. Moreover, we set
$J_1=-1$ and we
focus on weak and moderate IC $|J_{\perp}| \le 1$.

\subsection{Zero-temperature properties\label{sec:GS}}

\begin{figure}
%\centering \includegraphics[scale=0.65]{disprel_T0_J1x-1_J2x_FMandAFMcouplings_J1yJ1z011.eps} \protect\caption{
\centering \includegraphics[scale=0.65]{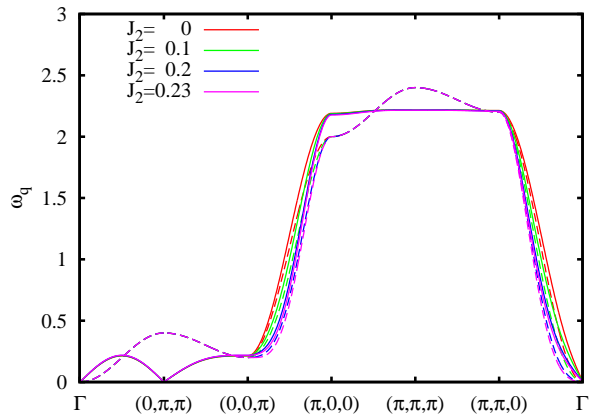} \protect\caption{(Color online)
Spin-wave dispersion $\omega_\mathbf{q}$ as a function of the wave vector $\mathbf{q}$ at zero temperature along several paths through the Brillouin Zone
(dashed lines: FM $J_{\perp}=-0.1$; solid
lines: AFM $J_{\perp}=0.1$).
Note that in the regions $\Gamma \ldots (0,0,\pi)$ and
$(\pi,0,0) \dots \Gamma$ all solid as well as all dashed lines coincide. 
}
\label{fig2} 
\end{figure}

\begin{figure}
%\centering \includegraphics[scale=0.65]{SpinWaveVelocity_x_T0_J1x-1_J2xvar_AFMcouplings_J1yJ1z_with_inset.eps} \protect\caption{
\centering \includegraphics[scale=0.65]{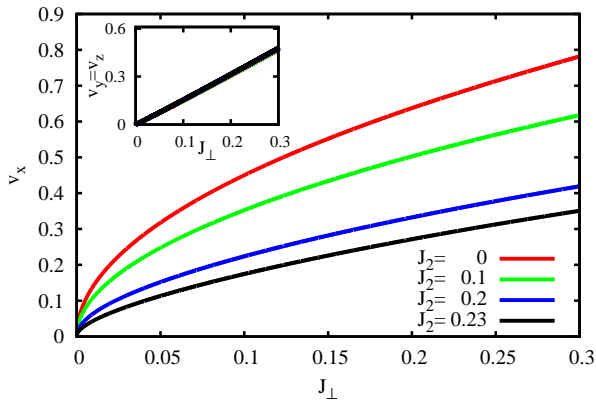} \protect\caption{(Color online)
GS spin-wave velocities $v_x$  (in-chain, main panel)  and
$v_y=v_z$  (inter-chain, inset) as a function of the AFM IC $J_{\perp}>0$ for different values of the frustrating NNN in-chain coupling $J_2$.
Note that the curves of the inter-chain velocities in the inset nearly 
coincide. 
}
\label{fig3} 
\end{figure}
\begin{figure}
%\centering \includegraphics[scale=0.65]{veloc_x_T0_J1yJ1z_J2x.eps} 
%\centering \includegraphics[scale=0.65]{veloc_and_rho_x_T0_J1yJ1z_J2x.eps}
\centering \includegraphics[scale=0.65]{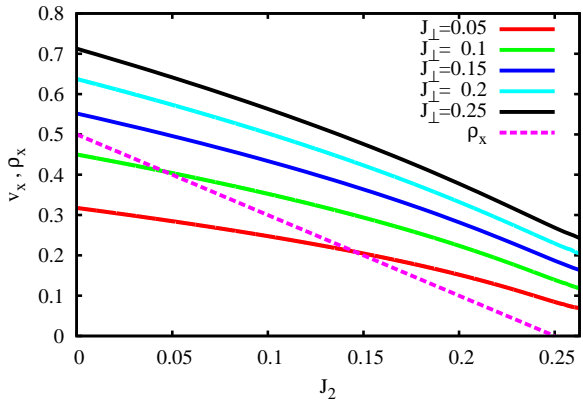}
\protect\caption{(Color online)
GS in-chain spin-wave velocity $v_x$ (solid lines, AFM $J_\perp$) as well as the 
in-chain spin stiffness $\rho_x$  
(dotted line, FM $J_\perp$) as a function of the frustration parameter  $J_2$ for different values of
the IC $J_{\perp}$.
Note that $\rho_x$  given by
$\rho_x=(\abs{J_1}-4J_2)/2$ is independent of  $J_{\perp}$. 
}
\label{fig4} 
\end{figure}

For ferromagnetic ICs $J_{\perp}$ and $0\leq
J_2<-J_1/4$ the GS is the fully polarized long-range ordered
ferromagnetic state, i.e., we have $\langle
S_\mathbf{0}S_\mathbf{R}\rangle=1/4$ and the total magnetization is $M=1/2$
(i.e., the condensation term is $C_{{\bf Q}^{FM}}=1/6$).
The corresponding spin-wave dispersion
$\omega_{\bf{q}}$ is shown in Fig.~\ref{fig2} (dashed lines) for
$J_{\perp}=-0.1$  and various values of $J_2$. Obviously, the influence of
$J_2$ on the general shape
of $\omega_{\bf{q}}$ is fairly weak. 
At the magnetic wave-vector ${\bf q}={\bf Q}^{FM}=\bf{0}$ ($\Gamma$ point) 
there is a quadratic dispersion (i.e.,$\omega_{q_i} \propto \rho_i q_i^2$, with $i=x,y,z$), that is typical for ferromagnets. 
The stiffness parameters, see also Eqs.~(\ref{stiffness_x}) and
(\ref{stiffness_y}), are given by $\rho_x=\abs{J_1+4J_2}/2$ (in-chain) and 
$\rho_\gamma=\abs{J_{\perp,\gamma}}/2$ ($\gamma=y,z$, inter-chain).

In the case of AFM ICs  $J_{\perp}>0$
 the GS is of quantum nature.  The corresponding magnetic
wave-vector is $Q^{AFM}=(0,\pi,\pi)$.
The  dispersion is linear for small values of $|\bf{q}|$, i.e., 
the low-lying excitations are determined by the spin-wave velocities
$v_x$ and $v_y=v_z$.
Again, 
 the influence of
$J_2$ on the general shape
of $\omega_{\bf{q}}$ is fairly weak, cf. the solid lines in Fig.~\ref{fig2}. 
Since several GS correlation functions enter the expressions for the
spin-wave velocities, cf. Eqs.~(\ref{velocity_x}) and  ~(\ref{velocity_y}), no
simple expressions can be given. However, it can be seen from these
equations that $v_x$, $v_y$, and $v_z$ are vanishing in the
limit $J_{\perp}\rightarrow0^+$ as expected. 
We show the spin-wave velocities in Figs.~\ref{fig3} and ~\ref{fig4}. 
Obviously,  the inter-chain spin-wave velocities  are almost linear functions in
$J_{\perp}$, i.e. $v_\gamma \sim
a J_{\perp}$, $\gamma=y,z$, and their dependence on           
the frustration parameter $J_2$ is weak, cf. the inset of
Fig.~\ref{fig3}.
The prefactor $a$ varies between $a=1.57$ at $J_2=0$ and $a=1.60$
at  $J_2=0.23$.
  On the other hand, the in-chain
spin-wave velocity $v_x$ exhibits a square-root like dependence on
$J_{\perp}$, cf. the main panel of
Fig.~\ref{fig3}. 
The influence of the in-chain frustration $J_2$ on  $v_x$ (relevant for AFM
$J_{\perp}$) and $\rho_x$ (relevant for FM
$J_{\perp}$) is shown in Fig.~\ref{fig4}.  

The main effect of the frustration consists  in a softening of the
long-wavelength
excitations,
i.e.  $v_x$  and  $\rho_x$
decrease with growing $J_2$, where  $v_x$ depends on $J_{\perp}$ and $\rho_x$
is independent of $J_{\perp}$.
However, in contrast to $\rho_x$ the spin-wave velocity  $v_x$  remains finite at
the transition point $J^c_2$, as it is known, e.g., for the square-lattice
$J_1-J_2$  model.\cite{nonlin_sigma,Schulz,ED40}

 Next we consider the magnetic order parameter $M$ for AFM IC,
which is related to the condensation
 term $C_\mathbf{Q}$ at
 the magnetic wave
 vector $\mathbf{Q}=\mathbf{Q}^{AFM}=(0,\pi,\pi)$, cf. Sec.~\ref{methods}. 
We show the dependence of $M$ on the IC in
Fig.~\ref{fig_4}. Starting from $M=1/2$ at  $J_{\perp}=0$ the
order parameter decreases  monotonously with increasing $J_\perp$
 indicating the role of
quantum fluctuations introduced to the system by AFM $J_{\perp}$. 
Moreover, it can be seen from Fig.~\ref{fig_4} that the larger $J_2$ the 
steeper the decrease of $M$ with growing $J_{\perp}$. 
A more explicit view on the influence of frustration $J_2$ on $M$ is
presented
in Fig.~\ref{fig_3}.
As can be expected already from  Fig.~\ref{fig_4}, we have a
monotonic decrease of the order parameter  with increasing $J_2$, i.e.
naturally frustration acts against magnetic ordering. 
The breakdown of the $Q^{AFM}=(0,\pi,\pi)$ long-range order at a critical
value  $J_2^c$
is indicated by a steep downturn of $M$.
A particular feature is the slight shift of the transition point  $J_2^c$
beyond the critical point of isolated chains, $J_2^c=1/4$, see
Fig.~\ref{fig_3}. Thus
we get $J_2^{c}\approx 0.256$ for $J_{\perp}=0.1$ and $J_2^{c}\approx 0.258$ for
$J_{\perp}=0.2$. Such a shift of $J_2^c$  to higher values was previously
also reported for the two-dimensional case, i.e. $J_{\perp,y} > 0$ and
$J_{\perp,z} = 0$, see Ref.~\onlinecite{zinke2009}.  

Finally we briefly discuss the  uniform static susceptibility $\chi_0$ for AFM
$J_\perp$, see Eq.~(\ref{eq_chi0}). Consistently,  $\chi_0$ diverges at
$J_{\perp}=0$. The inverse  uniform susceptibility, $1/\chi_0$, as a function of
$J_{\perp}$
is shown in the inset of
Fig.~\ref{fig_4}. Obviously,  $1/\chi_0$ is an almost linear function of
$J_{\perp}$, and the dependence on  
the frustration parameter $J_2$ is weak.
A fit according to  $\chi_0^{-1} =a J_{\perp}$ of the data shown in Fig.~\ref{fig_4} yields
$a = 12.25$,  $12.35$, $12.56$, and $12.69$ for
$J_2=0$,  $0.1$, $0.2$, and  $0.23$, respectively.

\begin{figure}
%\centering \includegraphics[scale=0.65]{Mag_T0_J1x-1_J2xvar_AFMcouplings_J1yJ1z_with_inset.eps} \protect\caption{
\centering \includegraphics[scale=0.65]{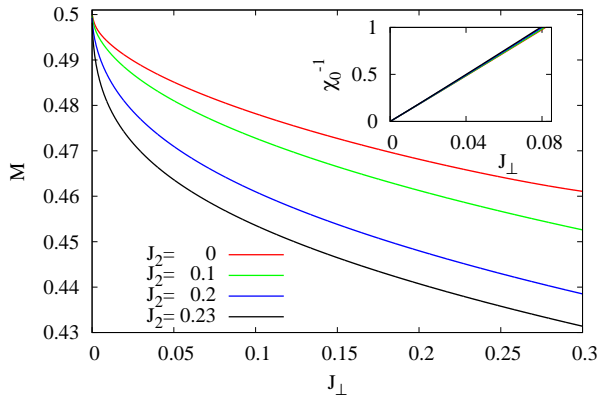} \protect\caption{(Color online)
GS magnetic order parameter $M$ (main panel) and inverse uniform
susceptibility $\chi_0^{-1}$ (inset) as a function of the AFM IC $J_{\perp}$ for different values of the 
frustrating NNN in-chain coupling $J_2$.
Note that the curves of the inverse uniform
susceptibility in the inset practically
coincide. 
}
\label{fig_4} 
\end{figure}

\begin{figure}
%\centering \includegraphics[scale=0.65]{Mag_T0_J1x-1_J1yJ1zvar_AFMcoupling_J2x.eps} \protect\caption{
\centering \includegraphics[scale=0.65]{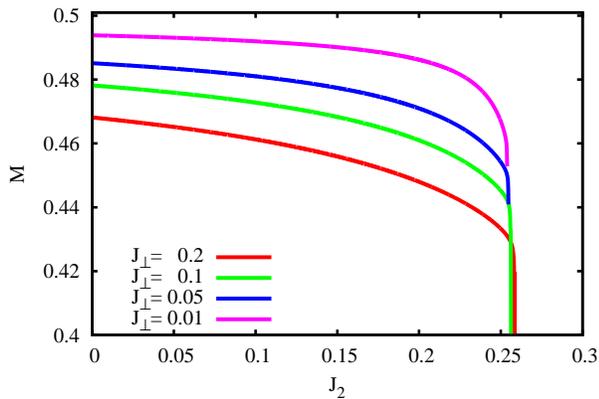} \protect\caption{(Color online)
GS magnetic order parameter $M$ as a 
function of the frustrating NNN in-chain coupling $J_2$ for different values of
AFM IC $J_\perp>0$.
}
\label{fig_3} 
\end{figure}

\subsection{Finite-temperature properties} 
\label{finite_T}
For the very existence of magnetic long-range order in an isotropic  Heisenberg
spin system at finite temperatures a
3D exchange  pattern is necessary,\cite{mermin66} i.e., finite ICs, $J_{\perp,y} \ne 0$ {\bf and} $J_{\perp,z} \ne 0$ are required.
Again  in this section we consider the special case of  $J_{\perp,y} =
J_{\perp,z}=J_{\perp}$. We mention that RGM data for the physical quantities
at arbitrary sets of  $J_2$, $J_{\perp,y}$ and $J_{\perp,z}$
are available upon request.

\subsubsection{Order parameters, critical temperatures and spin-spin correlation functions}
\label{subsec_Tc}

In Fig.~\ref{fig_M_T} we show some typical temperature profiles of the order
parameter calculated for $J_{\perp}= \pm 0.1$ and various values of
frustrating  $J_2$.  In accordance with previous studies on
quasi-two-dimensional unfrustrated spin systems \cite{oitmaa2004,junger2009}  we find that
for $J_2=0$ the transition temperature $T_c$  is larger
if AFM interactions are present. If $J_2>0$  the transition temperature is a result
of a subtle interplay of frustration $J_2$ and IC
$J_{\perp}$, since  these
parameters influence $T_c$ in  an opposite direction.
An illustration of the influence of $J_2$ and $J_{\perp}$ on $T_c$ is
provided in Figs.~\ref{fig_9} and \ref{fig_10}.
From Fig.~\ref{fig_9} (main panel) it is obvious that the slope of the $T_c(J_\perp)$ curve is
largest at $J_{\perp} \sim 0$. Moreover, following the trend
observed at $J_2=0$ we find that  $T_c$ for AFM $J_{\perp}
\gtrsim 0.1$ is larger than $T_c$ for corresponding FM IC irrespective of the
strength of frustration.
As we can see from Fig.~\ref{fig_10} (main panel) 
the reduction of $T_c$ due to
frustration is moderate as long as $J_2$ is not too close to the critical strength of frustration
$J_2^{c}$, where the FM GS ordering along the chains breaks
down.
Only as approaching $J_2^{c}$ there is a drastic downturn of $T_c$, cf. also
Ref.~\onlinecite{bcc_rgm_j1j2}.    

It is useful to compare the calculated critical temperatures with the
Curie-Weiss temperature $\Theta_{CW}$ given for the model at hand by
$\Theta_{CW}= -\frac{1}{2}(J_1+J_2+J_{\perp,y}+J_{\perp,z})$, where $J_1=-1$
(FM) and $J_2 \ge 0$
(AFM). 	
The absolute value of $\Theta_{CW}$ can be considered as a measure 
for the strength of the exchange interactions. Thus, in ordinary unfrustrated
3D magnets it determines the magnitude of the critical temperature $T_c$.  
The ratio  $f=|\Theta_{CW}/T_c|$ is often considered as the degree of frustration
see, e.g., Refs.~\onlinecite{Balents,Lang2016,Hallas2016}.
In conventional  3D ferro- and antiferromagnets
this ratio is of the order of unity, whereas  $f \gtrsim 5$ indicates a
suppression of magnetic ordering.
One may expect that also for unfrustrated or weakly frustrated  
quasi-2D (quasi-1D) systems in the limit of small
inter-layer (inter-chain) coupling the parameter $f$ can be large.   	
We show $f$ in the insets of Figs.~\ref{fig_9} and \ref{fig_10}.
Indeed from Fig.~\ref{fig_9} we notice that for $|J_\perp| < 0.05$ the ratio $f$
increases drastically. Thus, even for $J_2=0$ we find  $f>5$ at
$J_\perp<0.022$.       
The role of the frustrating coupling $J_2$ is illustrated in
Fig.~\ref{fig_10}.
It is obvious, that the influence of $J_2$ is weak in a wide range of $J_2$
values. Only as approaching the critical frustration $J_2^{c}$ there is
a tremendous increase of $f$ beyond $f>10$.     
We may conclude that the magnitude of the  frustration
parameter is a result of a subtle interplay of $J_\perp$ and $J_2$, and, a
large 
value  of $f$ does not unambiguously indicate frustration.

The order-disorder transition is also evident in the spin-spin correlation functions
$\langle {\bf S}_0{\bf S}_{\bf R}\rangle$, see Figs.~\ref{fig_corTTcJ20} and
\ref{fig_corTTcJ202}. Thus, for small $|J_\perp|$ the 
inter-chain correlations $\langle {\bf S}_0{\bf S}_{\bf{R}}\rangle$, ${\bf R}=(0,0,n)$,
become very small at $T>T_c$, whereas the correlations
along the chain direction, $\langle {\bf S}_0{\bf S}_{\bf{R}}\rangle$, ${\bf R}=(n,0,0)$, 
remain pretty large at $T\gtrsim T_c$ indicating the magnetic short-range
order along the chains in the paramagnetic phase. The effect of in-chain frustration $J_2$ is also
visible by comparing the green lines in  Figs.~\ref{fig_corTTcJ20} and
\ref{fig_corTTcJ202}.

\begin{figure}
%\centering \includegraphics[scale=0.65]{mag_J1x-1_J2xva_J1yJ1z0.1fmafm_T.eps} \protect\caption{
\centering \includegraphics[scale=0.65]{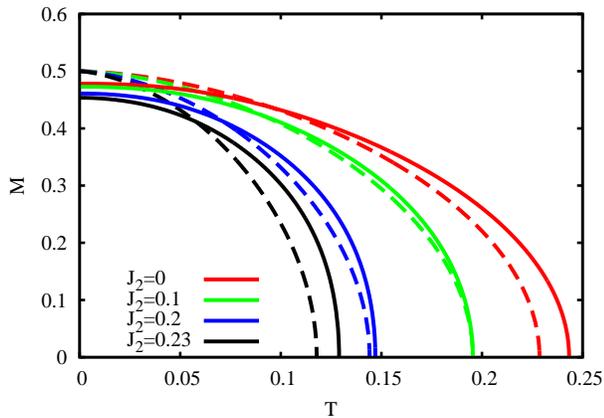} \protect\caption{(Color online)
Temperature dependence of the magnetic order parameter $M$ for
AFM $J_{\perp}=+ 0.1$ (solid lines) and FM $J_{\perp}=- 0.1$ (dashed
lines) and various values of the frustrating in-chain coupling $J_2$. 
}
\label{fig_M_T} 
\end{figure}

\begin{figure}
%\centering \includegraphics[scale=0.65]{Tcrit_J1x-1_J2x_FMandAFMcouplings_J1yJ1zvar_mult.eps}
\centering \includegraphics[scale=0.65]{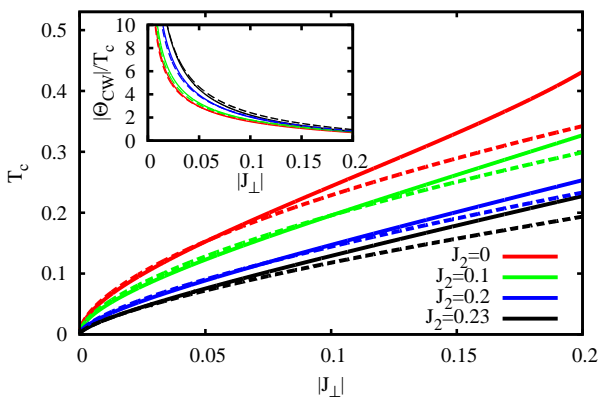} 
\protect\caption{(Color online)
Main panel: Critical temperature $T_c$ as a function of the IC $J_\perp$
(FM -
dashed; AFM - solid) for several values of the frustrating in-chain coupling $J_2>0$.
Inset: Ratio $f=|\Theta_{CW}/T_c|$ of the Curie-Weiss temperature $\Theta_{CW}=
-\frac{1}{2}(J_1+J_2+2J_{\perp})$ and the critical temperature $T_c$.
}
\label{fig_9} 
\end{figure}

\begin{figure}
%\centering \includegraphics[scale=0.65]{Tcrit_J1x-1_J1yJ1z_FMandAFMcouplings_J2xvar_mult.eps}
\centering \includegraphics[scale=0.65]{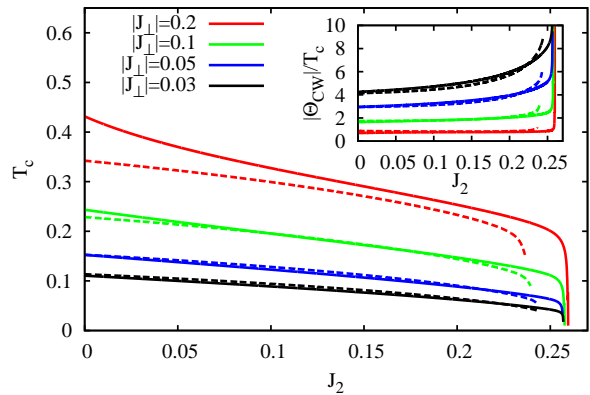} 
\protect\caption{(Color online)
Main panel: Critical temperature $T_c$ as a function of the frustrating in-chain coupling $J_2>0$ 
for several values of the IC $J_{\perp}$ 
(FM - dashed; AFM - solid).
Inset: Ratio $f=|\Theta_{CW}/T_c|$ of the Curie-Weiss temperature $\Theta_{CW}=
-\frac{1}{2}(J_1+J_2+2J_{\perp})$ and the critical temperature $T_c$.
}
\label{fig_10} 
\end{figure}

\begin{figure}
%\centering \includegraphics[scale=0.65]{corrfunc_TTc_J1x-1_J2x0var_couplings_J1yJ1z01.eps} \protect\caption{
\centering \includegraphics[scale=0.65]{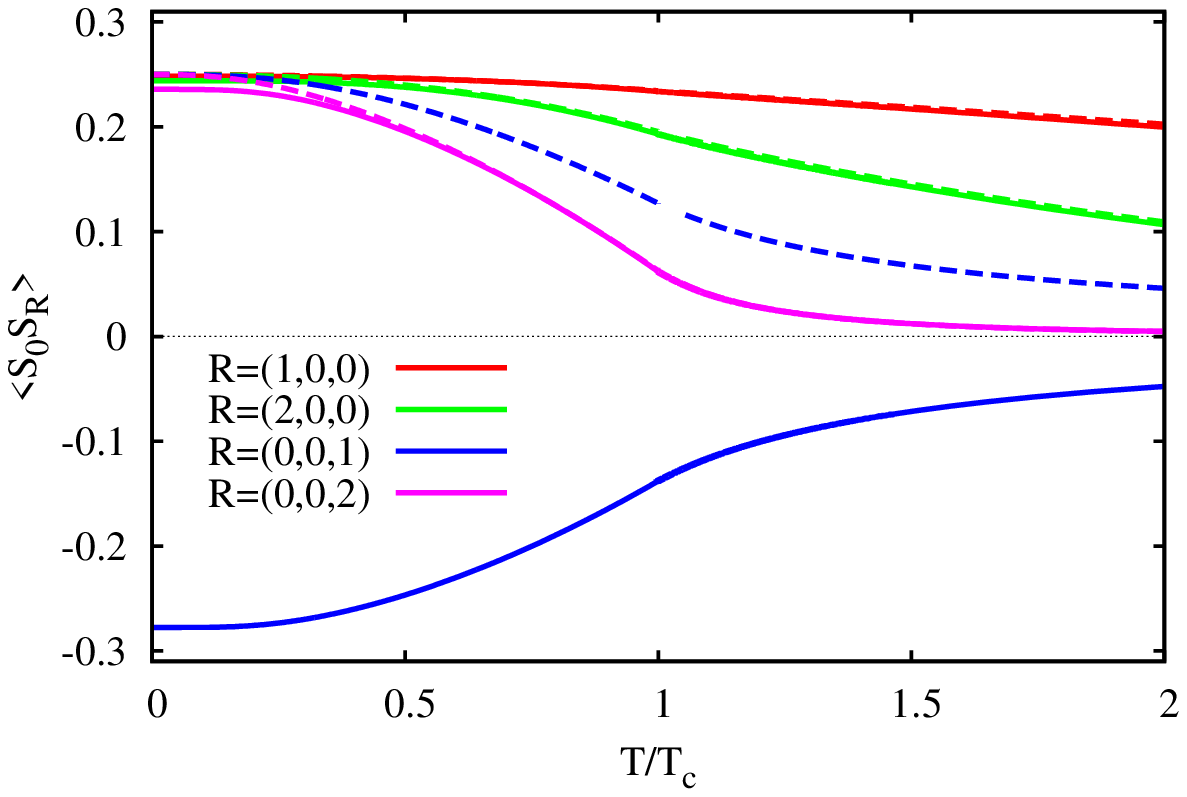} \protect\caption{(Color online)
Several spin-spin correlation functions as a function of the normalized temperature $T/T_c$ 
for the IC $\abs{J_{\perp}}=0.1$ 
(AFM - solid; FM - dashed) and for $J_2=0$. Note that the solid and dashed lines
are very close to each other (except for ${\bf R}=(0,0,1)$). 
}
\label{fig_corTTcJ20} 
\end{figure}

\begin{figure}
%\centering \includegraphics[scale=0.65]{corrfunc_TTc_J1x-1_J2x0.2var_couplings_J1yJ1z01.eps} \protect\caption{
\centering \includegraphics[scale=0.65]{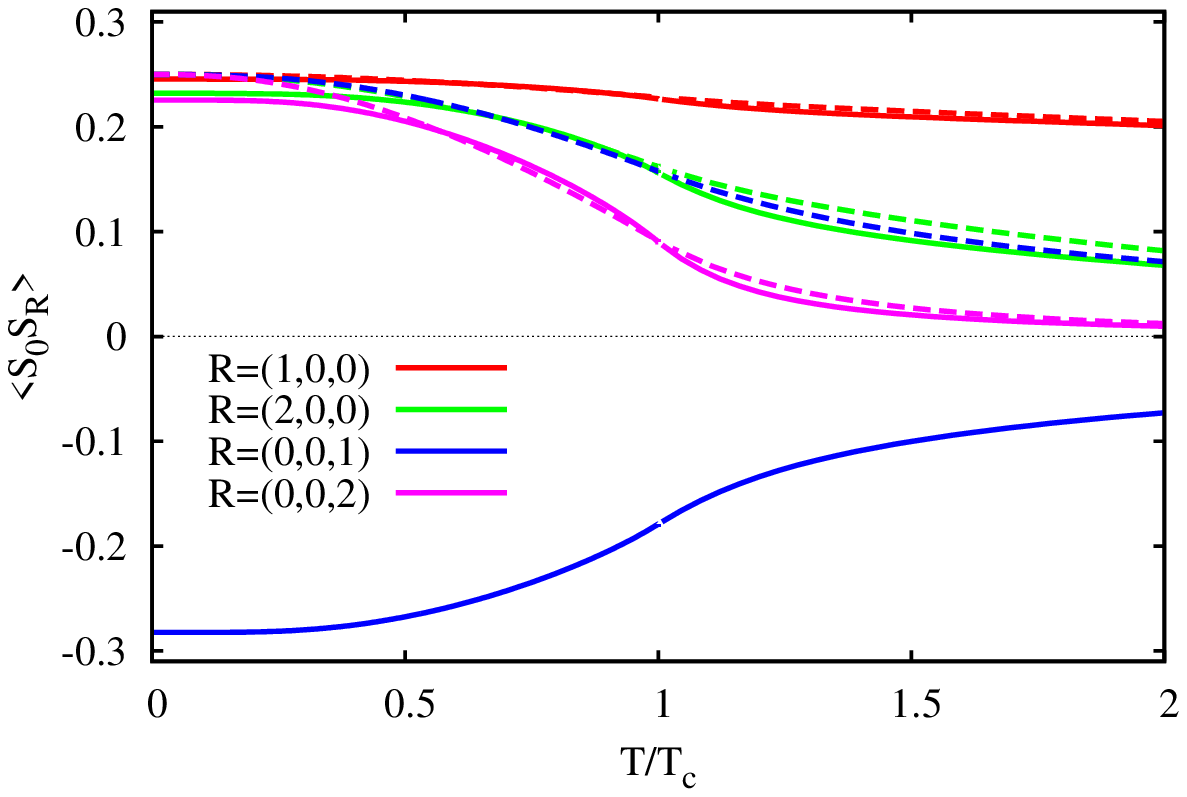} \protect\caption{(Color online)
Several spin-spin correlation functions as a function of the normalized temperature $T/T_c$ 
for the IC $\abs{J_{\perp}}=0.1$ 
(AFM solid; FM dashed) and for $J_2=0.2$.
Note that the solid and dashed lines
are very close to each other (except for ${\bf R}=(0,0,1)$). 
}
\label{fig_corTTcJ202} 
\end{figure}

\subsubsection{Correlation length and uniform static susceptibility}
\label{subsec_korrlength}

The correlation length, shown in Fig.~\ref{fig_11} for the unfrustrated
case, illustrates clearly  
the different behavior of the inter- and in-chain correlations, if $J_\perp$
is noticeably smaller 
than $J_1$. While the inter-chain correlation length drops down very rapidly
towards one lattice spacing for
$T \gtrsim T_c$, the   
in-chain correlation length remains quite large in a wider region above $T_c$
indicating the 1D nature of the magnetic behavior above the
transition.    
The role of the in-chain frustration on the correlation lengths becomes
evident by comparing Figs.~\ref{fig_11} and \ref{fig_12}. For strong
frustration $J_2=0.2$ used for the presentation in Fig.~\ref{fig_12}
the correlation lengths form a narrow bundle, i.e., the differences between
the in-chain and the inter-chain correlation lengths      
become much smaller compared to the case $J_2=0$, since the in-chain correlations on longer separations
are substantially diminished by frustration.    
\begin{figure}
%\centering \includegraphics[scale=0.65]{cor_length_J1x-1_J2x0_FMandAFM_J1yJ1z02.eps} 
\centering \includegraphics[scale=0.65]{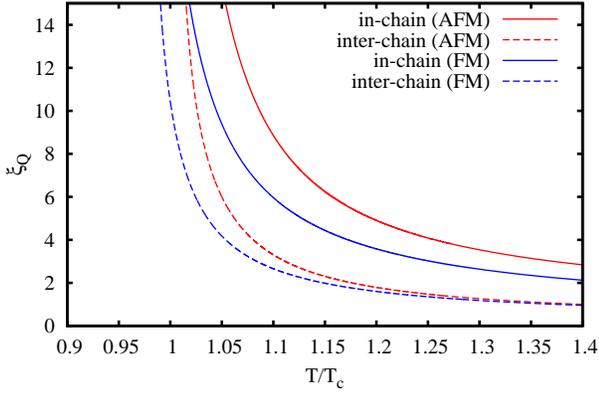} 
\protect\caption{(Color online)
Correlation length $\xi_\mathbf{Q}$ as a function of the normalized temperature $T/T_c$ for $J_2=0$ 
(FM $J_{\perp}=-0.2$ -- blue; AFM
$J_{\perp}=0.2$ -- red; in-chain correlation length
-- solid, 
inter-chain correlation length --
dashed).
}
\label{fig_11}
\end{figure}
\begin{figure}
%\centering \includegraphics[scale=0.65]{cor_length_J1x-1_J2x0.2_FMandAFM_J1yJ1z02.eps}
\centering \includegraphics[scale=0.65]{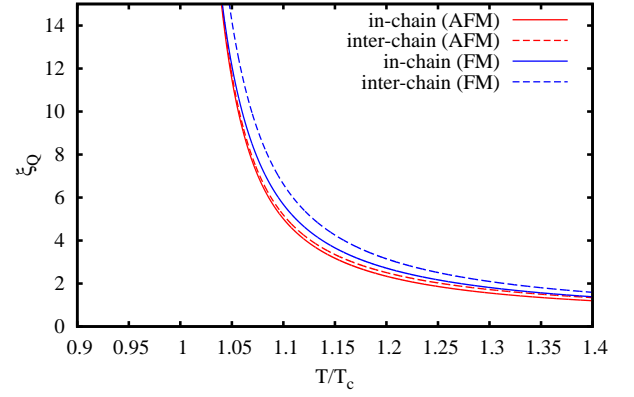}
\protect\caption{(Color online)
Correlation length $\xi_\mathbf{Q}$ as a function of the normalized temperature $T/T_c$ for $J_2=0.2$ 
(FM $J_{\perp}=-0.2$ -- blue; AFM
$J_{\perp}=+0.2$ -- red; in-chain correlation length
-- solid, 
inter-chain correlation length --
dashed).
}
\label{fig_12} 
\end{figure}

The temperature dependence of the susceptibility $\chi_0$ presented in Fig.~\ref{fig_12a} exhibits the
typical behavior of antiferromagnets (main panel) and ferromagnets
(left inset). The effect of frustration is evident for both FM and AFM
$J_\perp$.
For FM $J_\perp$ the overall shape of the curve is very similar for
different $J_2$. However, there is a noticeable shift towards higher
values of $T/T_c$ as increasing $J_2$.  For AFM $J_\perp$ the shape of $\chi_0(T)$ above
$T_c$ is
affected by $J_2$. For the IC of $J_\perp=0.1$ used in Fig.~\ref{fig_12a} the critical temperature $T_c$
is small and there
is a broad maximum in $\chi_0$ noticeably above $T_c$ related to the 
inter-chain antiferromagnetic correlations. By increasing $J_2$ the position of this maximum is shifted towards larger values
of $T/T_c$: it is at 
 $T/T_c=1.05$ for $J_2=0$ and at $T/T_c=1.23$ for $J_2=0.2$, see the right
inset in Fig.~\ref{fig_12a}.
On the other hand,
below $T_c$ the influence of $J_2$ on the $\chi_0(T/T_c)$ curves is very
weak. 
The influence of $J_\perp$ on the temperature profile of $\chi_0$ for AFM IC is
depicted in Fig.~\ref{fig_12b}.
Except the influence of the IC on the critical temperature discussed in
Sec.~\ref{subsec_Tc}
the strength of the AFM IC has also a strong influence on the magnitude  of the
uniform susceptibility at the transition point, $\chi_0(T_c)$, in case of weak
IC.
That is related to the behavior of $\chi_0$ in the limit $J_\perp \to 0+$,
where  we have $T_c \to 0$ and $\chi_0(T_c) \to \infty$.
Thus, as lowering $J_\perp$ from moderate values to zero, $\chi_0(T_c)$ increases
drastically. Below $T_c$ the AFM IC leads to a characteristic  downturn of
$\chi_0$,  cf. Fig.~\ref{fig_12b}.

\begin{figure}
%\centering \includegraphics[scale=0.65]{chi0_J1x-1_J2xvar_couplings_J1yJ1z01.eps}
\centering \includegraphics[scale=0.65]{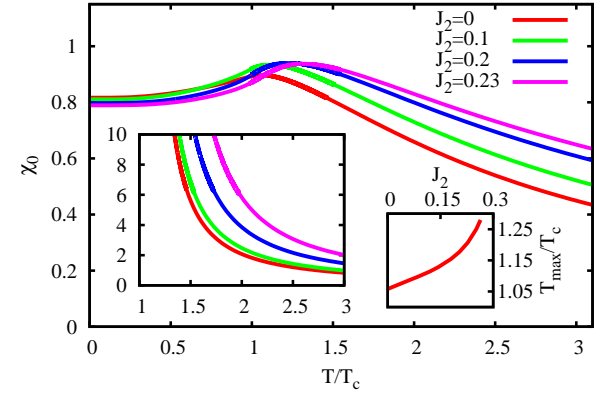} 
\protect\caption{(Color online)
Main panel: Uniform static susceptibility $\chi_0$ as a function of the 
normalized temperature $T/T_c$ for several values of the frustrating in-chain coupling $J_2$
and AFM $J_{\perp}=0.1$. Left inset: Uniform susceptibility $\chi_0$ as a function of the 
normalized temperature $T/T_c$ for several values of the frustrating in-chain coupling $J_2$
and FM $J_{\perp}=-0.1$. 
Right inset: Position of the maximum of the uniform susceptibility $\chi_0$,
$T_{\mbox{max}}/T_c$ as a function of $J_2$ for AFM $J_\perp=0.1$.
}
\label{fig_12a} 
\end{figure}

\begin{figure}
%\centering \includegraphics[scale=0.65]{chi0_J1x-1_J2x0_J1yJ1z_mult.eps}
\centering \includegraphics[scale=0.65]{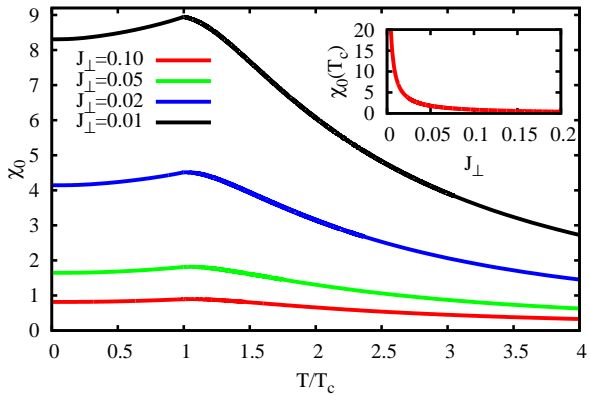}
\protect\caption{(Color online)
Main panel: Uniform susceptibility $\chi_0$ as a function of the 
normalized temperature $T/T_c$ for several values of the AFM IC $J_\perp$
and $J_2=0$. Inset: The value of the uniform susceptibility at the transition
temperature, $\chi_0(T_c)$ as a function of the 
AFM IC $J_\perp$ for $J_2=0$.
}
\label{fig_12b} 
\end{figure}

\subsubsection{Excitation spectrum  and specific heat}
\label{subsec_spinwave}

Finally we consider the temperature dependence of energetic quantities such as the specific heat
$C_V(T)$, the
 spin-wave velocities $v_\gamma$ (for AFM $J_\perp$) and the spin
stiffnesses $\rho_\gamma$ (for FM $J_\perp$), where $\gamma=x,y,z$. 
Let us start with a few remarks with respect to the comparison between
the RGM and the standard random-phase
approximation (RPA), see, e.g.,
Refs.~\onlinecite{bcc_rgm_j1j2,heisenbergferro2d,Tya67,du,froebrich2006,bcc_RPA_2014}.
The spin-wave 
excitation energies obtained within the framework of the RGM, see Eq.~(\ref{omega}), 
show a temperature renormalization 
that is
wavelength dependent and proportional to the correlation functions. Thus, as an example, the
existence of spin-wave excitations does not imply 
a finite magnetization. By contrast,  
within the RPA, the temperature renormalization of the
excitations
is independent of the wavelength and proportional to
the magnetization, see, e.g., Refs.~\onlinecite{Tya67,gasser}. % Gasser S. 289 
Moreover, 
the RPA fails in describing magnetic excitations and magnetic short-range order
for $T>T_c$, reflected, e.g., in the
specific heat.\cite{heisenbergferro2d,Tya67,gasser}

According to the above discussion on the temperature dependence of the
excitation spectrum, 
the RGM is appropriate to provide also information on the temperature dependence of
$v_\gamma$ and
$\rho_\gamma$ ($\gamma=x,y,z$), cf. Ref.~\onlinecite{bcc_rgm_j1j2}. 
We show the in-chain and inter-chain spin-wave velocities  (relevant for AFM
IC)  in
Figs.~\ref{fig_13} and \ref{fig_14}, respectively, and of the corresponding
stiffnesses (relevant for FM
IC) in Figs.~\ref{fig_15} and \ref{fig_16}, respectively.
Typically, the stiffness  and the spin-wave velocity decrease with
increasing temperature indicating a softening of spin excitations at $T>0$, cf.
Refs.~\onlinecite{Katanin2,Sun2006,Lovesey1977,bcc_rgm_j1j2,soft_swt1,soft_swt2,soft_swt3}.     
Interestingly, an opposite trend of the temperature influence on $v_x$ and  $\rho_x$
can emerge as increasing $J_2$ towards the transition point $J_2^c$.
That is in accordance with recent studies on other frustrated ferromagnets
\cite{bcc_rgm_j1j2,Katanin2}
and could therefore be interpreted as a signature of frustration in
(anti-)ferromagnets.
The  temperature dependence of $\rho_x$ at $J_2=0.23$, i.e.
very close to the transition point $J^c_2$, is somehow  special, since it is
first decreasing and then increasing with temperature.
 
As discussed already in Sec.~\ref{subsec_Tc}
the degree of frustration often is related to the ratio of the Curie-Weiss temperature $\Theta_{CW}$
and the transition temperature
$T_c$, i.e. to 
$f=|\Theta_{CW}/T_c|$.
We also mentioned in  Sec.~\ref{subsec_Tc} that
a large 
value  of $f$ does not unambiguously signalize frustration, since 
small values of $J_\perp$ also may lead to large values  of $f$  
even without any frustrating couplings.
Hence, the unusual temperature dependence of the spin-wave velocity and the
stiffness discussed above can be understood
as another criterion to detect frustration.

\begin{figure}
%\centering \includegraphics[scale=0.65]{AFM_mode_x_J1x-1_J2xvar_AFMcouplings_J1yJ1z01.eps} \protect\caption{
\centering \includegraphics[scale=0.65]{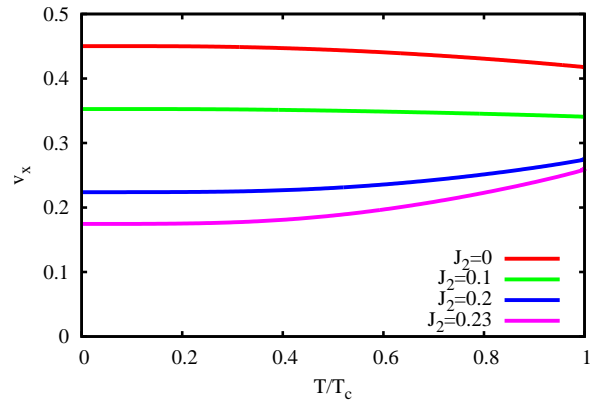} \protect\caption{(Color online)
In-chain spin-wave velocity $v_x$ as a function of the normalized temperature $T/T_c$
for AFM IC $J_{\perp}=0.1$ 
and for different values of the frustrating NNN in-chain coupling $J_2$.
}
\label{fig_13} 
\end{figure}

\begin{figure}
%\centering \includegraphics[scale=0.65]{AFM_mode_y_J1x-1_J2xvar_AFMcouplings_J1yJ1z01.eps} \protect\caption{
\centering \includegraphics[scale=0.65]{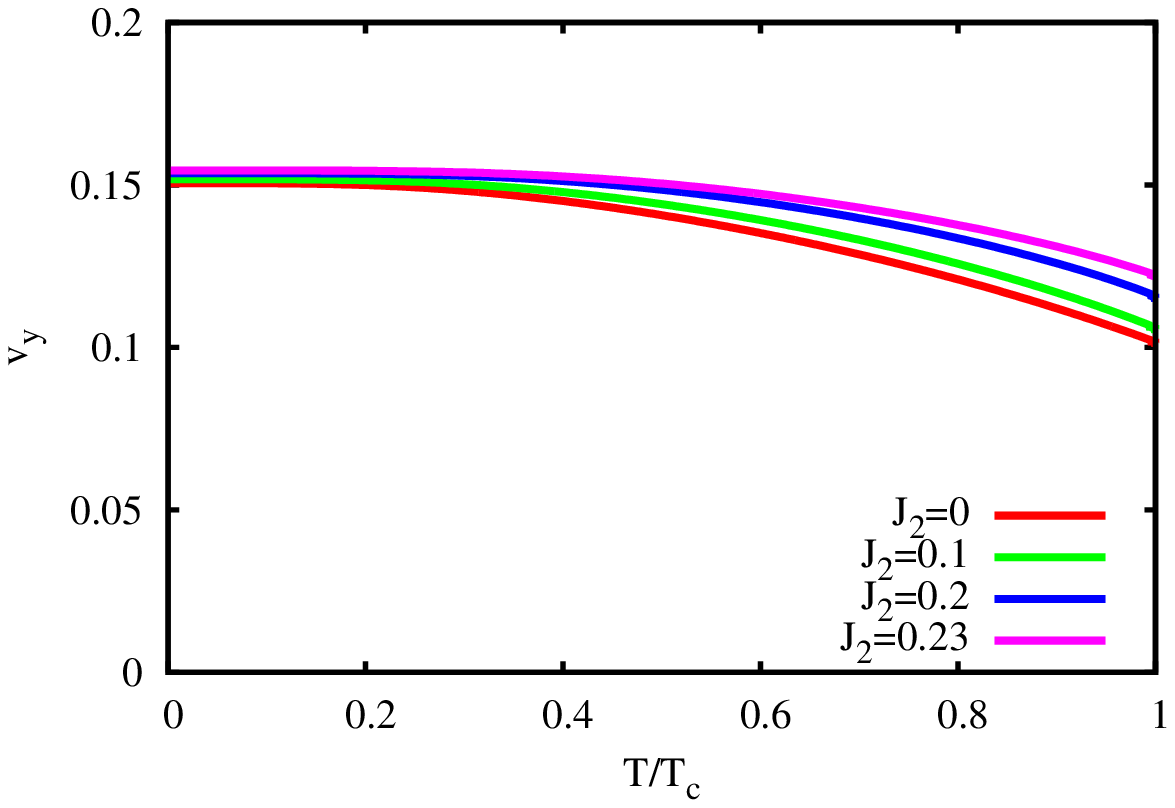} \protect\caption{(Color online)
Inter-chain spin-wave velocity $v_y=v_z$ as a function of the normalized temperature $T/T_c$
for  AFM IC $J_{\perp}=0.1$ 
and for different values of the frustrating NNN in-chain coupling $J_2$.
}
\label{fig_14} 
\end{figure}
\begin{figure}
%\centering \includegraphics[scale=0.65]{FM_mode_x2_J1x-1_J2xvar_FMcouplings_J1yJ1z01.eps} \protect\caption{
\centering \includegraphics[scale=0.65]{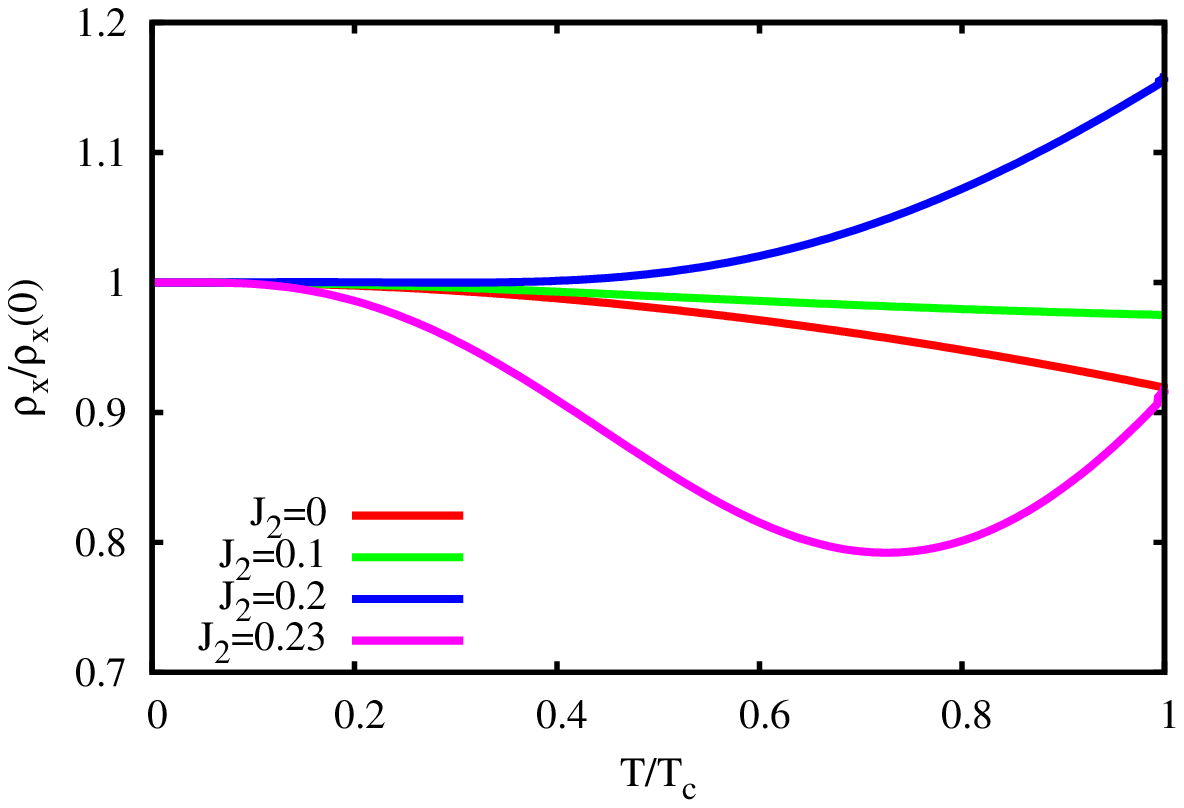} \protect\caption{(Color online)
In-chain spin stiffness $\rho_x$ scaled by its value at $T=0$ as a function of the normalized temperature $T/T_c$
for FM IC $J_{\perp}=-0.1$ 
for different values of the frustrating NNN in-chain coupling $J_2$.
}
\label{fig_15} 
\end{figure}

\begin{figure}
%\centering \includegraphics[scale=0.65]{FM_mode_y2_J1x-1_J2xvar_FMcouplings_J1yJ1z01.eps} \protect\caption{
\centering \includegraphics[scale=0.65]{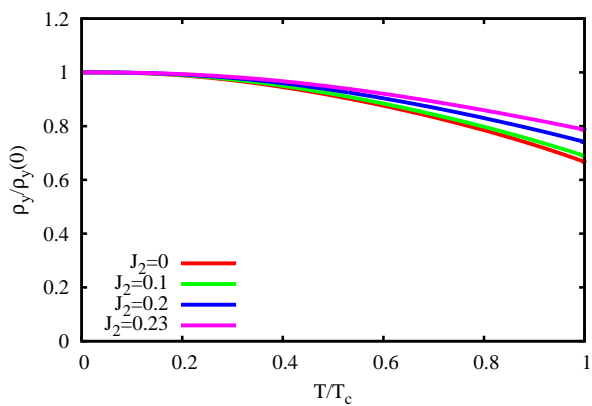} \protect\caption{(Color online)
Inter-chain spin stiffness $\rho_y=\rho_z$ scaled by its value at $T=0$ as a function of the normalized temperature $T/T_c$
for FM IC $J_{\perp}=-0.1$ 
for different values of the frustrating NNN in-chain coupling $J_2$.
}
\label{fig_16} 
\end{figure}

The temperature dependence of the specific heat $C_V$ is shown in
Fig.~\ref{fig_spec2}
for $J_2=0$ and two values of $J_\perp$.
The $C_V(T)$ curves  show the characteristic cusp-like 
behavior at the transition temperature
$T_c$ indicating the second-order phase transition. 
For very small values of $J_\perp$ above  the cusp a separate broad maximum emerges
which is related to the in-chain spin-spin correlations, i.e.,  the position of this
maximum  is mainly 
determined by the in-chain exchange parameters, cf.
Ref.~\onlinecite{RGMchainfrusferro}.     
\begin{figure}
%\centering \includegraphics[scale=0.65]{spec_heat_J1x-1_J2x0_and005_FMandAFM.eps} \protect\caption{
\centering \includegraphics[scale=0.65]{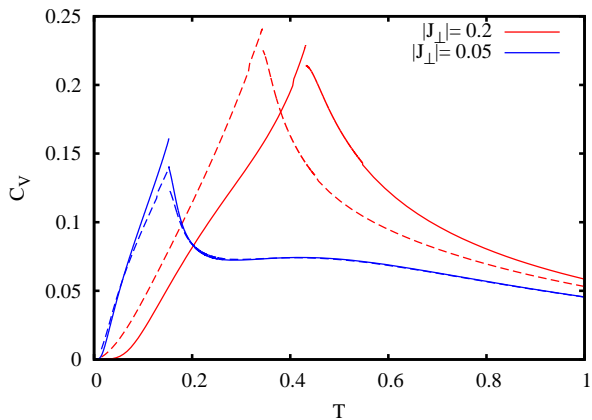} \protect\caption{(Color online)
Temperature dependence of the specific heat $C_V$ for various values of  
$J_{\perp}$ and $J_2=0$ (dashed lines: FM $J_{\perp}$; solid
lines: AFM $J_{\perp}$).
}
\label{fig_spec2} 
\end{figure}

\section{Summary}
\label{sec:sum} 
In our paper we investigate coupled frustrated
spin-$1/2$  $J_1$-$J_2$ Heisenberg chains with FM NN exchange $J_1$ and AFM NNN exchange
$J_2$. We consider 
FM as well as AFM inter-chain couplings (ICs)  $J_{\perp,y}$ and
$J_{\perp,z}$ corresponding to the axis perpendicular to the chain.
We focus on the regime of weak and moderate values of $J_2$, such that the
in-chain spin-spin correlations are predominantly FM.
 We use the rotation-invariant Green's function
method (RGM) to calculate thermodynamic quantities, such as the (sublattice)
magnetization  (magnetic
order parameter) $M$,
the critical temperature
$T_c$,  the correlation
functions
$\langle {\bf S}_0 {\bf S}_{\bf R} \rangle$,  the uniform static susceptibility
$\chi_0$, the correlation length
$\xi_{\bf Q}$, the specific heat $C_{V}$, the spin stiffnesses as well as the spin-wave
velocities.
The RGM  
goes one step beyond the random-phase
approximation (RPA).  As a result, several shortcomings of the RPA, see,
e.g.,
Refs.~\onlinecite{du,Tya67,gasser,froebrich2006,bcc_RPA_2014}, such as  
the artificial equality of the  critical temperatures $T_c$ for FM and
AFM couplings  or the failure in describing the paramagnetic
phase  at $T>T_c$, can be overcome. 
As approaching the ground-state transition point to the helical in-chain
phase at $J_2 \sim |J_1|/4$, the thermodynamic properties are strongly
influenced by the frustration. 
Thus, there is a drastic decrease of $T_c$ as $J_2 \to |J_1|/4$.   
Moreover, the temperature profile of
the in-chain spin stiffness $\rho_x$ (for FM IC) or the in-chain spin-wave
velocity (for AFM IC) may exhibit an increase with $T$
instead of the ordinary decrease.

The present investigations are focused on theoretical aspects, and we
consider  the simplest case of perpendicular ICs.  
Although, there are a few materials  corresponding
to perpendicular ICs, e.g.,
LiVCuO$_4$ and
Li(Na)Cu$_2$O$_2$ \cite{gippius2004,rosner,enderle2005,buettgen2007}, in real magnetic $J_1$-$J_2$ compounds typically  
the ICs are
more sophisticated than those we consider in our paper, see, e.g.,
Ref.~\onlinecite{prb2015}.\\

\begin{acknowledgements}
We thank  S.-L. Drechsler and O. Derzhko
for fruitful discussions. 
\end{acknowledgements}

\appendix*
\section{Analytical Expressions}

In this section we provide analytical expressions of the uniform
susceptibility $\chi_0$, the staggered susceptibility $\chi_{\mathbf{Q}=(0,\pi,\pi)}$, the spin-wave stiffnesses $\rho_i$ and the spin-wave velocities $v_i$ ($i=x,y,z$),
which enter the equations given in Sec.~\ref{methods}.\\

{\it Static
susceptibility:}
\begin{widetext}
\begin{eqnarray}
\label{eq_chi0}
&&\lim_{q_z\to0}\chi(q_x=0,q_y=0,q_z)=\chi_0^{(1)}\nonumber\\ 
&&\quad = -\frac{2 c_{001}}{-4 J_{1} p_{001}+4 J_{1} p_{101}-4 J_{\perp,y} p_{001}+4 J_{\perp,y} p_{011}-6 J_{\perp,z} p_{001}+2
   J_{\perp,z} p_{002}-4 J_{2} p_{001}+4 J_{2} p_{201}+J_{\perp,z}}  ,
\end{eqnarray}

\begin{eqnarray}
\label{eq_chi01}
&&\lim_{q_y\to 0}\chi(q_x=0,q_y,q_z=0)=\chi_0^{(2)}\nonumber\\ 
&&\quad =
 -\frac{2 c_{010}}{-4 J_{1} p_{010}+4 J_{1} p_{110}-6 J_{\perp,y} p_{010}+2 J_{\perp,y} p_{020}-4 J_{\perp,z} p_{010}+4
   J_{\perp,z} p_{011}-4 J_{2} p_{010}+4 J_{2} p_{210}+J_{\perp,y}} ,
\end{eqnarray}

\begin{eqnarray}
\label{eq_chi02}
&&\lim_{q_x\to 0}\chi(q_x,q_y=0,q_z=0)=\chi_0^{(3)} =
\frac{2 J_{1} c_{100}+8 J_{2} c_{200}}{\Delta^{(3)}_{\mathbf{0}}} ,
\end{eqnarray}

\begin{eqnarray}
\label{eq_Delta3}
\Delta^{(3)}_{\mathbf{0}} &= &J_{1}^2 (6 p_{100}-2 p_{200}-1)+2 J_{1} (2 J_{\perp,y}(p_{100}-p_{110})+2 J_{\perp,z} p_{100}-2 J_{\perp,z} p_{101}-3 J_{2} p_{100}+8 J_{2} p_{200}-5 J_{2} p_{300})\nonumber \\
   &+&4J_{2} (4 (J_{\perp,y} p_{200}-J_{\perp,y} p_{210}+J_{\perp,z} 
p_{200}-J_{\perp,z} p_{201})+J_{2} (6 p_{200}-2 p_{400}-1)),
\end{eqnarray}

\begin{eqnarray}
\label{eq_chi0pp}
 \chi_{(0,\pi,\pi)} &=& -\frac{2 (J_{\perp,y} c_{010}+J_{\perp,z} c_{001})}{\Delta_{(0,\pi,\pi)} } ,
\end{eqnarray}

\begin{eqnarray}
\label{eq_delta0pp}
 \Delta_{(0,\pi,\pi)} &=& 4 J_{\perp,y} (-J_{1} p_{010}+J_{1} p_{110}+J_{\perp,z}(p_{001}+p_{010}+2 p_{011})-J_{2} p_{010}+J_{2} p_{210})\nonumber \\
 &+&J_{\perp,z} (-4 J_{1} p_{001}+4 J_{1} p_{101}+2 J_{\perp,z}p_{001}+2 J_{\perp,z} p_{002}-4 J_{2} p_{001}+4 J_{2} p_{201}+J_{\perp,z})\nonumber \\
 &+&J_{\perp,y}^2 (2 p_{010}+2 p_{020}+1) .
\end{eqnarray}

{\it Spin-wave velocities:}
\begin{eqnarray}\label{velocity_x}
v_x^2 &=&J_{1}^2 \left(-3 p_{100}+p_{200}+\frac{1}{2}\right)\nonumber \\
&+&J_{1} (2 J_{\perp,y} (p_{110}-p_{100})-2 J_{\perp,z} p_{100}+2 J_{\perp,z}p_{101}+3 J_{2} p_{100}-8 J_{2} p_{200}+5 J_{2} p_{300})\nonumber \\
&+&2 J_{2} (-4 J_{\perp,y} p_{200}
+4 J_{\perp,y} p_{210}-4J_{\perp,z} p_{200}+4 J_{\perp,z} p_{201}-6 J_{2} p_{200}+2 J_{2} p_{400}+J_{2}),
\end{eqnarray}

\begin{eqnarray} \label{velocity_y}
2v_y^2/J_{\perp,y}  &=&-4 J_{1} p_{010}+4 J_{1} p_{110}-6 J_{\perp,y} p_{010}+2 J_{\perp,y} p_{020}-4 J_{\perp,z} p_{010}\nonumber \\
&+&4J_{\perp,z} p_{011}-4 J_{2} p_{010}+4 J_{2} p_{210}+J_{\perp,y} ,
\end{eqnarray}

\begin{eqnarray} \label{velocity_z}
2v_z^2/J_{\perp,z} &=& -4 J_{1} p_{001}+4 J_{1} p_{101}-4 J_{\perp,y} p_{001}+4 J_{\perp,y} p_{011}-6 J_{\perp,z} p_{001}\nonumber \\
&+&2J_{\perp,z} p_{002}-4 J_{2} p_{001}+4 J_{2} p_{201}+J_{\perp,z} .
\end{eqnarray}

{\it Spin  stiffnesses:}
\begin{eqnarray} \label{stiffness_x}
24\rho^2_x &=&J_{1}^2 (30 p_{100}-2 p_{200}-1)+16 J_{2} (4 (J_{\perp,y} p_{200}-J_{\perp,y} p_{210}+J_{\perp,z}p_{200}-J_{\perp,z} p_{201})+J_{2} (30 p_{200}-2 p_{400}-1))\nonumber \\
&+&2 J_{1} (2 J_{\perp,y} (p_{100}-p_{110})+2 J_{\perp,z} p_{100}
-2 J_{\perp,z} p_{101}+33J_{2} p_{100}+80 J_{2} p_{200}-17 J_{2} p_{300}),
\end{eqnarray}

\begin{eqnarray} \label{stiffness_y}
36\rho_{y}^2&=&-6J_{\perp,y}(J_{1}(p_{110}-p_{010})-J_{\perp,z}p_{010}
+J_{\perp,z}p_{011}-J_{2}p_{010}+J_{2}p_{210}) \nonumber\\
&-&J_{\perp,y}^{2}(3(p_{020}-15p_{010})+\frac{3}{2}) ,
\end{eqnarray}

\begin{eqnarray} \label{stiffness_z}
36\rho_{z}^{2}&=&-6J_{\perp,z}(J_{1}(p_{101}-p_{001})-J_{\perp,y}p_{001}
+J_{\perp,y}p_{011}-J_{2}p_{001}+J_{2}p_{201})\nonumber\\
&-&J_{\perp,z}^{2}(3(p_{002}-15p_{001})+\frac{3}{2}).
\end{eqnarray}
\end{widetext}

\end{document}